\documentclass[final]{IEEEtran}
\usepackage{amsmath}
\usepackage{color}
\usepackage{theorem}
\usepackage{subfigure}
\usepackage[lined,algonl,boxed,ruled,linesnumbered]{algorithm2e}
\usepackage{times}
\usepackage[latin1]{inputenc}
\usepackage{amsfonts}
\usepackage{psfrag}
\usepackage{amsbsy}
\usepackage{amssymb}
\usepackage[mathscr]{eucal}
\usepackage{latexsym}
\usepackage[T1]{fontenc}
\usepackage[dvips]{graphicx}
\usepackage{bm}
\usepackage{array}
\usepackage{showlabels}
\usepackage{dsfont}
\usepackage{mfirstuc}
\DeclareMathAlphabet{\mathpzc}{OT1}{pzc}{m}{it}

\usepackage{verbatim,mathrsfs, cite}
\usepackage[acronym,shortcuts]{glossaries}
\usepackage{soul}
\usepackage{tikz}
\usepackage{pgf}
\usetikzlibrary{dsp,chains}
\newacronym{ACK}{ACK}{acknowledge}
\newacronym{ARQ}{ARQ}{automatic repeat request}
\newacronym{AWGN}{AWGN}{additive white Gaussian noise}
\newacronym{BCC}{BCC}{broadcast channel with confidential messages}
\newacronym{BPSK}{BPSK}{binary phase shift keying}
\newacronym{CDF}{CDF}{cumulative distribution function}
\newacronym{CSI}{CSI}{channel state information}
\newacronym{DEC}{DEC}{decoder}
\newacronym{ENC}{ENC}{encoder}
\newacronym{HARQ}{HARQ}{hybrid automatic repeat request}
\newacronym{iid}{iid}{independent identically distributed}
\newacronym{IR-HARQ}{IR-HARQ}{incremental redundancy HARQ}
\newacronym{LDPC}{LDPC}{low density parity check}
\newacronym{MIMO}{MIMO}{multiple input multiple output}
\newacronym{MRC}{MRC}{maximal ratio combining}
\newacronym{NACK}{NACK}{not acknowledge}
\newacronym{PDF}{PDF}{probability density function}
\newacronym{PEG}{PEG}{progressive edge growth}
\newacronym{PMD}{PMD}{probability mass distribution}
\newacronym{RTD-HARQ}{RTD-HARQ}{repetition time diversity HARQ}
\newacronym{S-HARQ}{S-HARQ}{secure HARQ}
\newacronym{SNR}{SNR}{signal to noise ratio}
\newacronym{CER}{CER}{codeword error rate}
\newacronym{OFDM}{OFDM}{orthogonal frequency division multiplexing}
\newacronym{DF}{DF}{decode and forward}
\newacronym{AF}{AF}{amplify and forward}

{\theorembodyfont{\rmfamily}
   }
{\theorembodyfont{\rmfamily}
   }
{\theorembodyfont{\rmfamily}
   }
{\theorembodyfont{\rmfamily}
   }

{\theorembodyfont{\rmfamily}
   }
{\theorembodyfont{\rmfamily}
   }

\begin{document}

\title{Resource Allocation for Secure Gaussian Parallel Relay Channels with Finite-Length Coding \\ and Discrete Constellations}

\author{Linda Senigagliesi, Marco Baldi and Stefano Tomasin
\thanks{The material in this paper was presented in part at the IEEE Conference on Communications and Network Security (CNS 2015) -- Workshop on Physical-layer Methods for Wireless Security, Florence, Italy, Sep. 2015.}
\thanks{L. Senigagliesi and M. Baldi are with Dipartimento di Ingegneria dell'Informazione, Universit\`a Politecnica delle Marche, 60131 Ancona, Italy (e-mail: l.senigagliesi@pm.univpm.it, m.baldi@univpm.it).}
\thanks{S. Tomasin is with Department of Information Engineering, University of Padova, 35131 Padova, Italy (e-mail: tomasin@dei.unipd.it).}

}

\maketitle

\begin{abstract}
We investigate the transmission of a secret message from Alice to Bob in the presence of an eavesdropper (Eve) and many of decode-and-forward relay nodes.  Each link comprises a set of parallel channels, modeling for example an \acl{OFDM} transmission.  We consider the impact of discrete constellations and finite-length coding, defining an achievable secrecy rate under a constraint on the equivocation rate at Eve. Then we propose a power and channel allocation algorithm that maximizes the achievable secrecy rate by resorting to two coupled Gale-Shapley algorithms for  stable matching problem. We consider the scenarios of both full and partial \acl{CSI} at Alice. In the latter case, we only guarantee an {\em outage} secrecy rate, i.e., the rate of a message that remains secret with a given probability. Numerical results are provided for Rayleigh fading channels in terms of average outage secrecy rate, showing that practical schemes achieve a performance quite close to that of ideal ones.
\end{abstract}


\glsresetall

\section{Introduction}

Adding secrecy features to the physical layer is an active and promising research area \cite{Bloch-book}, that complements traditional computational security approaches. 
Indeed, a proper coding scheme can  prevent an eavesdropper Eve from getting information on a message exchanged between the two legitimate users Alice and Bob \cite{Harrison2013}.

In this paper we expand the results of \cite{Tomasin2015} on resource allocation for confidential communications over the Gaussian parallel relay channels, by including the more practical constraints of  finite-length coding and discrete constellations. We first derive the achievable secrecy rate of this scheme under the assumption of full \ac{CSI} by Alice and the relay nodes. Then in order to consider the impact of discrete constellations and finite-length coding we define an achievable secrecy rate under a constraint on the equivocation rate at Eve. Using an approximated formula of the achievable secrecy rate we derive the optimal power allocation for point-to-point confidential transmission. By exploiting the power and rate adaptation algorithm for the parallel relay channels of \cite{Tomasin2015}, we obtain a resource allocation algorithm coupling two Gale and Shapley algorithms to allocate resources over the parallel relay channels. We also consider the partial \ac{CSI} scenario, in which Alice does not know the gains of her channels to Eve, while their statistics are known. In this case we only guarantee an {\em outage} secrecy rate, i.e., the rate of a message that remains secret with a given probability. We show that the algorithm derived for full \ac{CSI} can be easily adapted to the partial \ac{CSI} scenario. Numerical results are provided, showing the merit of the proposed solution.

\subsection{Related Works}

\begin{figure*}
\centering
\scalebox{1}{
\begin{tikzpicture}
	\node[circle,draw] (tx) {Alice};
	
	% Alice-relay channel	
	\node[coordinate, right= of tx] (in2) {};
	\node[coordinate, above= of in2, yshift=1em] (in1) {};
	\node[coordinate, below= of in2, yshift=-2em] (in3) {};
	\node[dspmixer, right= of in1] (ch1) {};
	\node[coordinate, right= of in2] (ch2) {};
	\node[dspmixer, right= of in3] (ch3) {};
	\node[coordinate, below= of ch1,yshift=1em] (h1) {};
	\node[coordinate, below= of ch2,yshift=1em] (h2) {} ;
	\node[coordinate, below= of ch3,yshift=1em] (h3) {};
	\node[dspadder,right= of ch1] (a1) {};		
	\node[coordinate,right= of ch2] (a2) {};		
	\node[dspadder,right= of ch3] (a3) {};
	\node[coordinate, below= of a1,yshift=1em] (n1) {};
	\node[coordinate, below= of a2,yshift=1em] (n2) {};		
	\node[coordinate, below= of a3,yshift=1em] (n3) {};
	\node[circle,draw,right= of a1,minimum size=.8cm] (rx1) {$r_1$ };		
	\node[coordinate,right= of a2,] (rx2) {};	
	\node[coordinate,below= of rx2,yshift=3em] (punti) {}; 	
	\node[circle,draw,right= of a3,minimum size=.8cm] (rx3) {$r_N$};
	
	% relay-Bob channel
	\node[dspmixer, right= of rx1,xshift=8em] (ch1b) {};
	\node[coordinate, right= of rx2,xshift=8em] (ch2b) {};
	\node[dspmixer, right= of rx3,xshift=8em] (ch3b) {};
	\node[coordinate,right= of ch1b,xshift=.4em] (c1) {};
	\node[coordinate,right= of ch2b,xshift=.7em] (c2) {};
	\node[coordinate,right= of ch3b,xshift=1em] (c3) {};
	\node[coordinate, below= of ch1b,yshift=1em] (h1b) {};
	\node[coordinate, below= of ch2b,yshift=1em] (h2b) {} ;
	\node[coordinate, below= of ch3b,yshift=1em] (h3b) {};
	\node[dspadder, right= of c2] (rxb) {};
	\node[dspadder,right= of rxb] (ab) {};		
	\node[coordinate, below= of ab,yshift=1em] (nb) {};		
	\node[circle,draw,right= of ab] (Bo) {Bob};
	\node[coordinate, above= of rxb, yshift=.55em] (in1b) {};
	\node[coordinate, below= of rxb, yshift=-1.55em] (in3b) {};
	
	% Alice-Eve channel
	\node[dspmixer, below= of ch3, yshift=-3.7em] (chae) {};
	\node[dspadder,right= of chae] (ae1) {};		
	\node[coordinate, below= of ae1,yshift=0.5em] (ne1) {};	
	\node[circle,draw,right= of ae1,minimum size=1cm] (Ev) {Eve};

	% relay-Eve channel
	\node[coordinate, right= of rx1, xshift=4.9em] (tch1e) {};		
	\node[coordinate, right= of rx2, xshift=1.7em] (tch2e) {};	
	\node[coordinate, right= of rx2, xshift=4.9em, yshift=0.3em] (salto1) {};					
	\node[coordinate, right= of rx2, xshift=4.9em, yshift=-0.3em] (salto12) {};					
	\node[coordinate, right= of rx3, xshift=4.9em, yshift=0.3em] (salto13) {};					
	\node[coordinate, right= of rx3, xshift=4.9em, yshift=-0.3em] (salto14) {};					

	\node[coordinate, right= of rx3, xshift=1.7em, yshift=0.3em] (salto2) {};					
	\node[coordinate, right= of rx3, xshift=1.7em, yshift=-0.3em] (salto22) {};					
	
	\node[coordinate, right= of rx3, xshift=-1.5em] (tch3e) {};				
	\node[dspmixer, below= of ch3b, xshift=-3.5em, yshift=1.5em] (ch1e) {};
	\node[coordinate, left= of ch1e,xshift =0.5em] (ch2e) {};
	\node[dspmixer, left= of ch2e,xshift =-0.4em] (ch3e) {};
	\node[coordinate, right= of ch1e, xshift=.4em] (c1e) {};
	\node[coordinate, right= of ch2e, xshift=.7em] (c2e) {};
	\node[coordinate, right= of ch3e, xshift=1em] (c3e) {};
	\node[coordinate, left= of ch1e,xshift=1.2em] (h1e) {};
	\%node[coordinate, right= of ch2e,xshift=-1.2em] (h2e) {};
	\node[coordinate, left= of ch3e,xshift=1.2em] (h3e) {};
	\node[circle, right= of ch3e,xshift=-1cm] (punti2) {$\ldots$};
	\node[coordinate, below= of chae,yshift=0.5em] (hae) {};
	\node[dspadder, below= of ch2e,yshift=0.5em] (rxe) {};
	\node[coordinate, below= of ch3e,yshift=0.5em] (rxe1) {};
	\node[coordinate, below= of ch1e,yshift=0.5em] (rxe3) {};
	\node[dspadder,right= of Ev,xshift=1.1em] (ae) {};		
	\node[coordinate, below= of ae,yshift=0.5em] (ne) {};		
	\node[coordinate, above= of rxe, yshift=.55em] (in1e) {};
	\node[coordinate, right= of rxe] (in3e) {};	
	\node[coordinate, below= of in3 ] (inmiddle) {};			
	\node[coordinate, below= of inmiddle,yshift=-1.7em ] (inE) {};	

	% connections
	\draw[dspline] (tx) -- node[midway,above] {$\bm{P} \bm{1}_K$} (in2); 
	\draw[dspline] (in2) -- (in1); 	
	\draw[dspline] (in2) -- (in3); 	
	\draw[dspline] (in2) -- (inE); 		
	\draw[dspconn] (inE) -- (chae); 			
	\draw[dspconn] (in1) -- (ch1); 	
	%\draw[dspconn] (in2) -- (ch2); 	
	\draw[dspconn] (in3) -- (ch3); 		
	\draw[dspconn] (ch1) -- (a1); 		
	%\draw[dspconn] (ch2) -- (a2); 				
	\draw[dspconn] (ch3) -- (a3); 					
	\draw[dspconn] (a1) --  (rx1); 		
	%\draw[dspconn] (a2) --  (rx2); 				
	\draw[dspconn] (a3) --  (rx3); 				
	\draw[dspconn] (rxb) -- (ab);	
	\draw[dspconn] (ab) -- node[midway,above] {}(Bo); 		
	\draw[dspconn] (rx1) -- node[left,above] {$\bar{\bm{P}}_1$}  (ch1b); 		
	%\draw[dspconn] (rx2) -- node[left,above] {$\bar{\mathcal X}_2$}  (ch2b); 		
	\draw[dspconn] (rx3) -- node[left,above] {$\bar{\bm{P}}_N$}  (ch3b); 		
 	\draw[dspconn] (nb) -- node[midway,right] {$\sigma_n^2$}(ab);
 	\draw[dspconn] (n1) -- node[midway,right] {$\sigma_n^2$}(a1);
 	%\draw[dspconn] (n2) --  node[midway,right] {$\bm{w}_2$}(a2);
 	\draw[dspconn] (n3) --  node[midway,right] {$\sigma_n^2$}(a3);
	\draw[dspconn]  (h1) --  node[midway,right] {$\bm{H}_1$}(ch1); 	
	%\draw[dspconn] (h2) -- node[midway,right] {$\bm{H}_2$}(ch2); 	
	\draw[dspconn] (h3) -- node[midway,right] {$\bm{H}_N$}(ch3); 	
	\draw[dspconn] (h1b) -- node[midway,right] {$\bar{\bm{H}}_1$} (ch1b); 	
	%\draw[dspconn] (h2b) -- node[midway,right] {$\bar{\bm{H}}_2$} (ch2b); 	
	\draw[dspconn] (h3b) -- node[midway,right] {$\bar{\bm{H}}_N$} (ch3b); 	

	\draw[dspline] (tch1e)--(salto13);
	%\draw (salto12) arc (-90:90:.3em);
	%\draw[dspline] (salto1)--(salto13);	
	\draw (salto14) arc (-90:90:.3em);	
	\draw[dspconn] (salto14)--(ch1e);		
	%\draw[dspline] (tch2e)--(salto2);
	%\draw (salto22) arc (-90:90:.3em);
	%\draw[dspconn] (salto22)--(ch2e);		
	
	\draw[dspconn] (tch3e)--(ch3e);

	\draw[dspline] (ch1b) -- (in1b); 			
	\draw[dspline] (ch3b) -- (in3b); 				
	\draw[dspconn] (in1b) -- (rxb); 	
	\draw[dspconn] (in3b) -- (rxb); 		
	%\draw[dspconn] (ch2b) -- (rxb);
	%\draw[dspline] (ch1e) -- (in1e); 			
	%\draw[dspline] (ch3e) -- (in3e); 	
	\draw[dspline] (ch3e) -- (rxe1); 				
	\draw[dspline] (ch1e) -- (rxe3); 						
	\draw[dspconn] (rxe3) -- (rxe); 	
	\draw[dspconn] (rxe1) -- (rxe); 		
	%\draw[dspconn] (ch2e) -- (rxe);
 	\draw[dspconn] (ne) -- node[midway,right]{$\sigma_n^2$} (ae);	
%	\draw[dspconn] (rxe) -- (Ev);	
	\draw[dspconn] (h1e) -- node[midway,below] {$\bar{\bm{G}}_1$} (ch1e); 	
	%\draw[dspconn] (h2e) -- node[midway,below] {$\bar{\bm{G}}_2$} (ch2e); 	
	\draw[dspconn] (h3e) -- node[midway,below] {$\bar{\bm{G}}_N$} (ch3e);
	\draw[dspconn] (hae) -- node[midway,right]{$\bm{G}$}(chae);	
	\draw[dspconn] (ne1) -- node[midway,right]{$\sigma_n^2$}(ae1);		
	\draw[dspconn] (chae) -- (ae1);		
	\draw[dspconn] (ae1) -- node[midway,above] {} (Ev);		
	\draw[dspconn] (ae) -- node[midway,above] {} (Ev);		
	\draw[dspconn] (rxe) -- (ae);		

\end{tikzpicture}}
\caption{Power flow of the relay parallel channels with $N$ relays, $r_1, \ldots r_N$. Mixers $\otimes$ and adders $\uplus$ represent element-wise multiplication and addition  of vectors, respectively. }
\label{figmod}
\end{figure*}
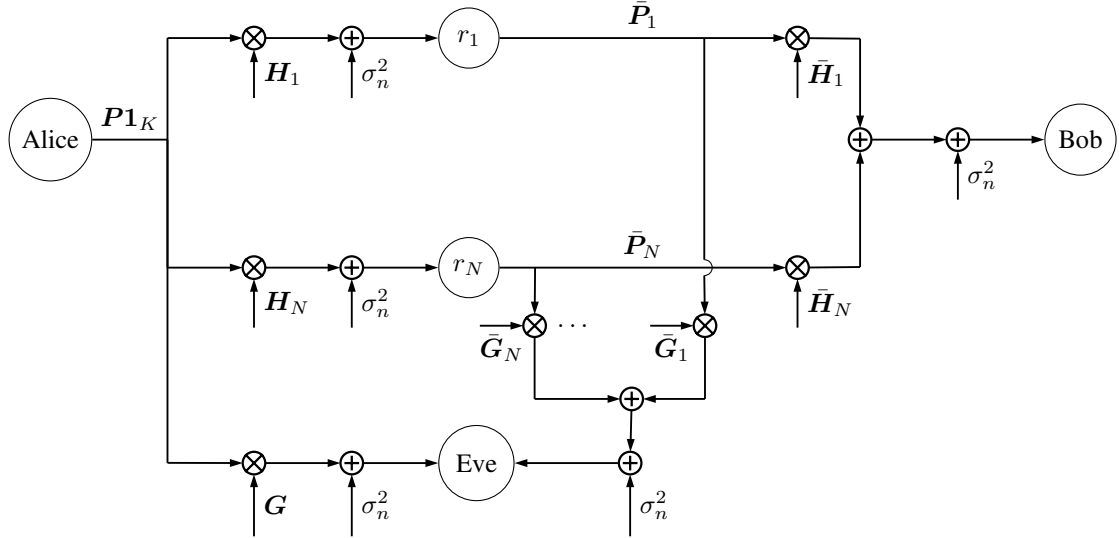

The physical layer security of messages transmitted over parallel channels with the assistance of trusted relays has already been addressed in the literature. Most works consider that relays can either forward the message or generate a noise signal to jam Eve. For links comprising a single channel, early works have addressed the relay selection problem  \cite{Wang-2012, Shen-nov12, Luo-icassp11, Shen-arxiv}, while various combinations of message forwarding and jamming are considered in \cite{Ding-jul12, Dong-mar10,Li-oct11} with multiple antennas nodes. In \cite{Bassily-jun12} multiple relays either jam or forward noise, i.e., they transmit random codewords from a globally known codebook, that hurts more Eve than Bob.

We focus on links comprising parallel channels. For this scenario, in \cite{Awan2012} rate-equivocation regions are derived by considering one relay only and assuming full \ac{CSI}. 
In \cite{Ng-oct11} \ac{OFDM} is considered with a single  relay, and Eve is equipped with multiple antennas under partial \ac{CSI}: subcarriers, powers and rates are optimized to maximize the average secrecy outage capacity. In \cite{Ng-glc11} the downlink of a cellular system is considered - where the multi-antenna base station  performs both beamforming and jamming against a single multi-antenna eavesdropper, and an outage problem is formulated under partial \ac{CSI}. 
The scenario is extended in \cite{Ng-cwit11}, where multiple relays operate in \ac{DF} mode and still an outage approach is considered.  In \cite{Ma-12} a single relay with parallel channels is considered, which performs cooperative jamming against Eve, under full \ac{CSI}. When the single relay performs \ac{DF}, resource optimization has been considered in \cite{Jeong-11}. More comprehensive results, considering also the direct transmission from Alice to Bob are obtained in \cite{Jeong-nov11}. Resource allocation for transmission over parallel channels assisted by \ac{DF} relays without secrecy features has also been widely studied. Bit loading  \cite{Gui-2008} and power and rate allocation \cite{Vandendorpe-2009} have been investigated, while the availability of multiple relays transmitting on a single sub-carrier is studied in \cite{Vandendorpe-2011}, with efficient greedy algorithms provided in \cite{Bakanoglu2011}. The resource allocation for parallel channels with secrecy outage constraint has been considered in \cite{Tomasin-icc14} and \cite{BaldiTomasinTIFS}. 

Recently, optimal resource allocation for security purposes under different conditions has gained the attention of several authors.
In \cite{Chen2016}, an optimization framework is proposed for two-hop communications that jointly optimizes source and  relay powers, and transmission time in each hop, with the goal of maximizing the secrecy outage capacity in a massive \ac{MIMO} scenario.
In \cite{Fang2017}, optimal power allocation and pricing strategies are determined using a Stackelberg game model in order to maximize the players' utilities, under both of perfect and imperfect \ac{CSI} assumptions and in the presence of multiple eavesdroppers.
An optimal power strategy to maximize the achievable secrecy rates in wireless multi-hop DF relay networks with a power constraint is studied in \cite{Lee2016}, under the assumption of global \ac{CSI}, and an iterative cooperative beamformer design is also proposed. The work is extended in \cite{Lee2018} to the case of full-duplex relays, with cooperative beamforming to null out the signal at multiple eavesdroppers. In \cite{Zhang2017} a heuristic resource allocation iterative algorithm is presented, based on the proximal theory that maximizes the secure capacity of device-to-device communications in heterogeneous networks. Joint source-relay power optimization  in a dual-hop communication using duality theory is performed in \cite{Aman2018}, with the aim of maximizing the overall secrecy rate, under individual power constraints and using an high SNR approximation. In \cite{Abedi2018}, a robust resource allocation framework is proposed in the presence of an active eavesdropper, assuming that both the legitimate receiver and the eavesdropper are full-duplex: the receiver sends jamming signals against the eavesdroppers, without the need for external helpers, and uncertain \ac{CSI} on the links between the eavesdropper and the legitimate receivers is considered.
Optimization algorithms for null space beamforming with full \ac{CSI} have been proposed in \cite{Obeed2018}.

\subsection{Contribution}

With respect to the previously described state of the art, the main contributions of this paper can be
summarized as follows.
\begin{itemize}
\item No existing work considers the role of finite-length codes and discrete constellations on the secrecy rate. Motivated by this, we provide a formulation of the secrecy rate under practical constraints and compare it with the achievable rate in ideal conditions. We consider both perfect and partial \ac{CSI} under outage constraints.
\item By proposing an approximated expression for the secrecy rate under practical conditions we optimize the link-level parallel channel power allocation generalizing the solution of the ideal transmission scenario.
\item Extending the approach in \cite{Tomasin2015}, we maximize the secrecy rate by resorting to an iterative algorithm based on the Gale and Shapley theory for the stable matching problem.
\item Through numerical examples we show that it is possible to achieve acceptable performance from the secrecy rate standpoint even using short codes and constellations with a small alphabet.
\end{itemize}

The rest of the paper is organized as follows. Section \ref{sec:system_model} outlines the system model for secret message transmission over parallel Gaussian relay channels. The achievable secrecy rates under full \ac{CSI} are computed in Section \ref{sec:secrecy_rate}, where we also compute the outage secrecy rate. In Section \ref{sec:pointtopoint} an algorithm for resource allocation of a secure point-to-point transmission over parallel channels is obtained, which is used then in Section \ref{sec:MRallocSec} for the resource allocation in a relay network. Numerical results of the proposed solution are presented in Section \ref{sec:num_results}, before some conclusions are drawn in Section \ref{sec:conclusions}.

\textit{Notation:} Vectors and matrices are written in bold letters. We denote the base-2 and natural-basis logarithm by $\log$ and $\ln$, respectively. We indicate the positive part of a real quantity $x$ as $\left[x \right]^+ {=} \max \lbrace x ; 0 \rbrace $. $\mathbb{E}[X]$ denotes the expectation of the random variable $X$, $\mathbb P[\cdot]$ is the probability operator, and $^T$ denotes the matrix transpose operator. The entropy is denoted as $\mathbb H(\cdot)$, while the mutual information is denoted as $\mathbb I(\cdot; \cdot)$.

\section{System Model}
\label{sec:system_model}

We consider a communication system to transmit a confidential message $\mathcal M$ from Alice to Bob through $N$ trusted cooperating relays.
Any link between a pair of devices is constituted by a set of $K$ parallel \ac{AWGN} channels. Eve is an eavesdropping device that overhears communications originated from both  Alice and the relays. No direct link between Alice and Bob is available, and all devices operate in half-duplex mode. Therefore  the message transmission comprises two phases: 
\begin{enumerate}
\item[1)] Alice transmits to the relays, and 
\item[2)] the relays transmit to Bob. 
\end{enumerate}
We also assume that in phase 2 at most one relay  transmits on channel $k$  and that the two phases have the same duration.
%(i.e., $N$ transmissions occur in each phase).  

Fig. \ref{figmod} shows the power flow of the considered scenario. We indicate with $P_{n, k}$ the transmit power of Alice on channel $k$ to relay $n$ in phase 1, while $\bar{P}_{n, k}$ is the transmit power of  relay $n$ on channel $k$ in phase 2.  
The $N \times K$ matrix $\bm{P}$ ($\bar{\bm{P}}$) collects all transmit powers, having $P_{n, k}$ ($\bar{P}_{n, k}$) at entry $n,k$. In Fig. \ref{figmod}, $\bm{P}\bm{1}_K$ denotes the $N$-size column vector of transmit powers for each relay, with $\bm{1}_K$ being the $K$-size column vector of all ones. We consider power constraints for both Alice and the relays, i.e.,
\begin{subequations}
\label{powerconst}
\begin{equation}
\sum_{k=1}^K P_{n,k} \leq P_{\rm tot,1}, \quad n=1, \ldots, N\,.
\end{equation}
\begin{equation}
\sum_{k=1}^K \bar{P}_{n,k} \leq P_{\rm tot,2}\,, \quad n=1, \ldots, N\,.
\end{equation}
\end{subequations}

The power constraint per relay in phase (1) simplifies the power allocation in this phase and still provides an upper bound on the total transmit power from the source, that can not exceed $NP_{\rm tot, 1}$.

The link from Alice to relay $n$ is represented by the $K$-size column vector $\bm{H}_n = [H_{n,1}, \ldots, H_{n,K}]^T$ containing the gains for each channel. The power of the data signal received by relay $n$ on channel $k$ is therefore $H_{n,k} P_{n,k}$. Similarly, the vector $\bar{\bm{H}}_n$ denotes the power gains of the link between relay $n$ and Bob and $\bar{H}_{n,k} \bar{P}_{n,k}$ is the power of the data signal received by Bob from relay $n$ on channel $k$. For links to Eve, $\bm{G}$ is the vector of power gains of the signal coming from Alice, while $\bar{\bm{G}}_n$ is the power gain vector of the signal coming from relay $n$. 

The noise is assumed to be \ac{iid}, with zero mean and unitary ($\sigma_n^2=1$ in Fig. \ref{figmod}) variance for all channels. Therefore, the \ac{SNR} at relay $n$ for a transmission from Alice on channel $k$ is $H_{n,k} P_{n,k}$, and similarly for a transmission from relay $n$ on channel $k$ the \ac{SNR} at Bob in phase 2 is $\bar{H}_{n,k}\bar{P}_{n,k}$.

\section{Achievable Secrecy Rate}
\label{sec:secrecy_rate}

We consider a per-channel encoding, i.e., Alice splits $\mathcal M$ into $K$ messages $\mathcal M_k$, $k=1, \ldots, K$, each of which is separately encoded and transmitted on a channel. In \cite{BaldiTomasinTIFS} an in-depth analysis of this coding strategy is provided, showing that it performs similarly to the scheme with joint coding across channels, while being simpler to design. Therefore, each relay in general receives only a subset of the entire message bits. In the second phase again each relay splits the received secret bits into groups, which are separately encoded and transmitted on a different channel, among those assigned to the relay. 

In both phases, secrecy is achieved through classical wiretap coding \cite{Bloch-book}, based on adding random bits to the secret message and encoding the resulting block with capacity achieving codes. The  {\em weak} secrecy rate of a point-to-point transmission is the rate of a message $\mathcal M$ that \cite{Bloch-book}: $i)$ is correctly decoded by Bob and $ii)$ has a rate of mutual information with the signal received by Eve $\mathcal Z$ that is vanishing for infinite codewords, i.e., 
\begin{equation}
\lim_{l \rightarrow \infty} \frac{1}{l} {\mathbb I}(\mathcal Z ; \mathcal M) = 0\,,
\label{vanishInfo}
\end{equation}
where $l$ is the message length in bits.
Due to the per-channel encoding, the achievable weak secrecy rate is the sum of the achievable secrecy rates on each used channel.
Let $R_{n,k}$ be the secrecy rate on channel $k$, intended for relay $n$ in phase 1, and $\bar{R}_{n,k}$ the secrecy rate on channel $k$ transmitted by relay $n$ in phase 2. The achievable secrecy rate between Alice and Bob is the minimum between the secrecy rates in both phases, i.e.,
\begin{equation}
\label{rtot}
R_{\rm tot}(\bm{P}, \bar{\bm{P}}) = \frac{1}{2} \sum_{n=1}^N \min\left\{\sum_{k=1}^K R_{n,k}(P_{n,k}), \sum_{k=1}^K \bar{R}_{n,k}(P_{n,k})\right\}\,,
\end{equation}
where the factor $1/2$ is due to the two phases of the same duration, and we have highlighted the dependence of the achievable rates on the transmit powers. 

Since we assume that Alice is transmitting to a single relay per channel we also have
\begin{equation}
R_{n^*,k}(P_{n^*,k}) > 0 \rightarrow R_{n,k}(P_{n,k}) = 0\,,\; n \neq n^*\,, 
\label{ratecons1}
\end{equation}
and since we assume that at most one relay is transmitting in any channel in phase 2 we also have
\begin{equation}
\bar{R}_{n^*,k}(\bar{P}_{n^*,k}) > 0 \rightarrow \bar{R}_{n,k}(\bar{P}_{n,k}) = 0\,,\; n \neq n^*\,.
\label{ratecons2}
\end{equation}

In the following we derive the achievable secrecy rates, when full \ac{CSI} is available at Alice, taking into consideration infinite and finite-length coding, continuous and discrete modulation formats. Then with discuss the $\epsilon$-outage achievable secrecy rates when Alice has only a partial \ac{CSI}, i.e., she knows only the statistics of her channels to Eve.

\subsection{Infinite-length coding with Gaussian Constellations}

When infinite-length coding and Gaussian constellations are used, perfect secrecy, i.e., no information leakage to Eve, can be achieved \cite{Bloch-book}. In this case,  the achievable secrecy rate can be written as 
\begin{equation}
R_{n,k}(P_{n,k}) = C(P_{n,k}H_{n,k}) - C(P_{n,k}G_{n,k})\,,
\label{seccapa}
\end{equation}
where $C(x) = \log(1+x)$. Similar expressions are obtained for $\bar{R}_{n,k}(\bar{P}_{n,k})$ where $P_{n,k}$,  $H_{n,k}$, and $G_{n,k}$ are replaced by $\bar{P}_{n,k}$, $\bar{H}_{n,k}$ and $\bar{G}_{n,k}$, respectively.

\subsection{Finite-length Coding with Gaussian Constellations}
\label{sec:FinInf}

A first limitation to the achievable secrecy rates introduced by practical systems is related to the use of codes working on finite-length blocks of symbols. 
For simplicity and adherence to practical systems, we consider deterministic coding, according to which each $l$-bit block of data is univocally mapped into a codeword $\mathcal{C}_{n,k}$. 
This is opposed to either random or coset coding, which are often invoked in the literature for this kind of systems, but yield further issues (e.g., concerning the generation of randomness).
In our setting, weak secrecy cannot be guaranteed and Eve can get some information on the secret message\footnote{Indeed, the definition of weak secrecy (\ref{vanishInfo}) entails a limit to infinity of the message length that can not be used in finite-codewords schemes.}. Moreover, the decodability condition at Bob cannot be guaranteed, and we will consider a non-zero \ac{CER} $\kappa$.

In order to measure the information leakage to Eve we resort to the equivocation rate, i.e., Eve's uncertainty about the message after observing the transmitted codeword (through her channel). For relay $n$ transmitting on channel $k$ and using codewords of $m$ symbols, the equivocation rate per symbol is 
\begin{equation}
\rho_{n,k}(P_{n,k}, R_{n,k}(P_{n,k})) = \frac{1}{2 m}{\mathbb H}({\mathcal M_k}|{\mathcal Z}_{n,k})\,,
\end{equation}
\begin{equation}
\bar{\rho}_{n,k}(\bar{P}_{n,k}, \bar{R}_{n,k}(\bar{P}_{n,k}))= \frac{1}{2 m}{\mathbb H}({\mathcal M_k}|\bar{\mathcal Z}_{n,k})\,,
\end{equation}
where the factor 2 comes from the fact that we have two phases of the same duration. We have that 
\begin{equation}
0 \leq \rho_{n,k}(P_{n,k}, R_{n,k}(P_{n,k})) \leq R_{n,k}(P_{n,k})\,,
\end{equation}
\begin{equation}
0 \leq \bar{\rho}_{n,k}(\bar{P}_{n,k}, \bar{R}_{n,k}(\bar{P}_{n,k})) \leq \bar{R}_{n,k}(\bar{P}_{n,k}),
\end{equation}
where the upper bound is achieved with infinitely-long codewords ($m\rightarrow \infty$). We will consider that the transmission system is secure if  
\begin{equation}
\frac{\rho_{n,k}(P_{n,k}, R_{n,k}(P_{n,k}))}{R_{n,k}(P_{n,k})} \geq \theta\,, \quad \frac{\bar{\rho}_{n,k}(\bar{P}_{n,k}, \bar{R}_{n,k}(\bar{P}_{n,k}))}{\bar{R}_{n,k}(\bar{P}_{n,k})} \geq \theta\,,
\label{eqrho}
\end{equation}
where  $\theta \in (0, 1]$ is a suitably defined parameter that limits the gap with respect to weak secrecy conditions with infinite-length coding. The {\em achievable secrecy rates} for finite-length coding are therefore the maximum rates satisfying condition (\ref{eqrho}), i.e.,
\begin{subequations}
\label{delRout1}
\begin{equation}
R_{n,k}(P_{n,k}) = \max_r r 
\end{equation}
s.t.
\begin{equation}
\frac{\rho_{n,k}(P_{n,k}, r)}{r} \geq \theta\,,
\end{equation}
\begin{equation}
{\mathbb P}[\mathcal M_n \neq \hat{\mathcal M}_n] = \kappa\,.
\end{equation}
\end{subequations}
A similar problem can be written for phase 2,  for a given allocated power $\bar{P}_{n,k}$, i.e.,
\begin{subequations}
\begin{equation}
\bar{R}_{n,k}(\bar{P}_{n,k}) = \max_r r 
\end{equation}
s.t.
\begin{equation}
\frac{\bar{\rho}_{n,k}(\bar{P}_{n,k}, r)}{r} \geq \theta\,,
\end{equation}
\begin{equation}
{\mathbb P}[\mathcal M_n \neq \hat{\mathcal M}_n] = \kappa\,.
\end{equation}
\label{delRout2}
\end{subequations}
In the following we focus on problem (\ref{delRout1}) as problem (\ref{delRout2}) is analogous.

For the computation of the equivocation rate we resort to a lower bound. By the definition of entropy and mutual information, we have that Eve's equivocation rate can be rewritten as
\begin{equation}
\rho_{n,k}(P_{n,k}, R_{n,k}(P_{n,k})) =  \frac{1}{m} \left[{\mathbb H}(\mathcal{C}_{n,k}) - {\mathbb I}(\mathcal{C}_{n,k}; \mathcal{Z}_{n,k})\right]\,,
\label{eq:equiv}
\end{equation}
where $\mathcal Z_{n,k}$ is the signal received by Eve in phase 1. By the definition of spectral efficiency as upper bound to the mutual information, we also have 
\begin{equation}
{\mathbb I}({\mathcal C}_{n,k}; {\mathcal{Z}}_{n,k})<  mC(P_{n,k}G_{n,k})\,.
\end{equation}
On the other hand, the entropy of ${\mathcal C}_{n,k}$ depends on the code rate, which in turns determines the (non null) \ac{CER}  $\kappa$ at relay $n$, due to the use of finite-length coding. A Gaussian approximation  on the code rate as a function of the \ac{CER} for finite-length coding is provided by \cite{Polyanskiy-13}, that can be written as
\begin{equation}
{\mathbb H}({\mathcal C}_{n,k}) \simeq m  \left[  C(P_{n,k}H_{n,k}) - \frac{\log e}{\sqrt{2m}} Q^{-1}(\kappa) \right]^+\,,
\label{eq:polyanskiy}
\end{equation}
where  $[x]^+ = x$ for $x \geq 0$ and 0 otherwise, and $Q(\cdot)$ is the complementary cumulative distribution function of the standard Gaussian variable. Therefore, we have the following approximation on Eve's equivocation rate
\begin{equation}
\begin{split}
\rho_{n,k}(P_{n,k}, &R_{n,k}(P_{n,k})) \simeq \xi{n,k}(P_{n,k}) \triangleq \\
&\left[ C(H_{n,k}P_{n,k}) - \frac{\log e}{\sqrt{2m}} Q^{-1}(\kappa) - C(G_{n,k}P_{n,k}) \right]^+.
\end{split}
\label{eq:eqbound}
\end{equation}

By replacing $\rho_{n,k}(P_{n,k}, R_{n,k}(P_{n,k}))$ with its approximated lower bound $\xi_{n,k}(P_{n,k})$ (and similarly for phase-2 equivocation rates) in  problems (\ref{delRout1}) and (\ref{delRout2}), we obtain the approximated achievable rates in the two phases; the solution can be obtained by numerical methods.

We consider a fitting of $R_{n,k}(P_{n,k})$ solution of (\ref{delRout1}) by the linear combination of logarithms of the powers, in order to ease resource allocation, i.e.,
\begin{equation}
\begin{split}
R_{n,k}&(P_{n,k}) \simeq  \alpha_1 + \alpha_2 \log(1 + \alpha_3 H_{n,k} P_{n,k}) - \\
& \alpha_4 \log(1 + \alpha_5 H_{n,k} P_{n,k}) - \left[   \alpha_6 \log(1 + \alpha_7 G_{n,k} P_{n,k}) - \right.\\
&  \left. \alpha_8 \log(1 + \alpha_9 G_{n,k} P_{n,k}) \right]\,.
\end{split}
\label{fitfunc}
\end{equation}
Note that (\ref{fitfunc}) directly models the achievable secrecy rate rather the equivocation rate, and the parameters $\alpha_i$ are chosen at solution of problem (\ref{delRout1}). By this formulation the secrecy rates with ideal conditions can be seen as a sub-case of (\ref{fitfunc}) with $\alpha_i = 1$ for $i=2,3,4,5$ and $\alpha_i=0$ otherwise.

\begin{figure}
\includegraphics[width=1\hsize]{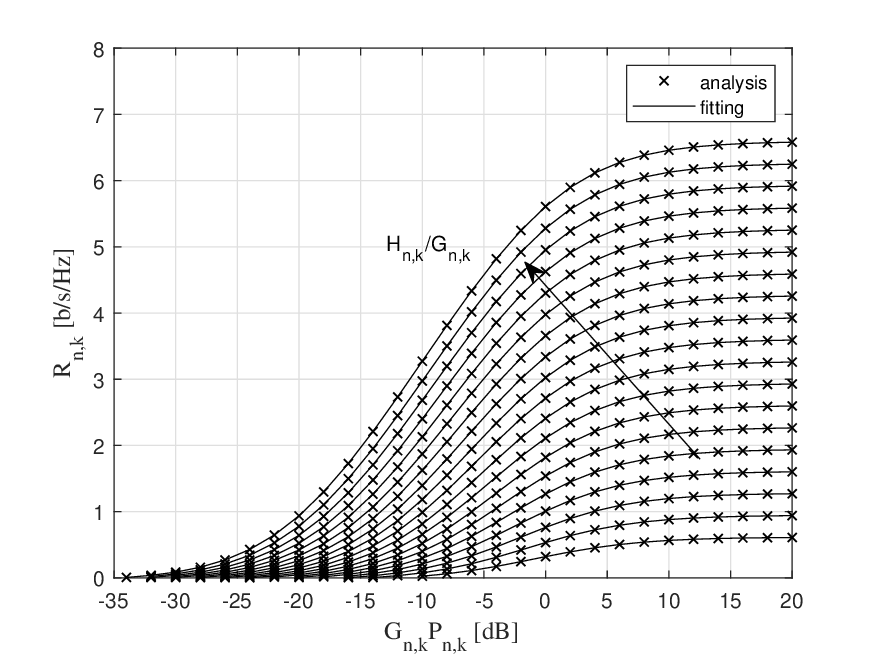}
\caption{$R_{n,k}(P_{n,k})$ as a function of $G_{n,k}P_{n,k}$ for values of $H_{n,k}/G_{n,k}$ between 2 dB and 20 dB with a step of 1 dB, and results obtained with the fitting function (\ref{fitfunc}).}
\label{fitting}
\end{figure}

Fig. \ref{fitting} shows $R_{n,k}(P_{n,k})$ as a function of $G_{n,k}P_{n,k}$ for values of $H_{n,k}/G_{n,k}$ between 2 dB and 20 dB with a step of 1 dB, and results obtained by the fitting function (\ref{fitfunc})  with $\kappa = 10^{-3}$ and $m = 4,096$. We observe a good agreement of the fitting function with $R_{n,k}(P_{n,k})$, especially at low rates, and high values of $G_{n,k} P_{n,k}$, with a slight overestimation of the rate for intermediate values of $G_{n,k} P_{n,k}$ for high $H_{n,k}/G_{n,k}$ ratios.

\subsection{Infinite-length Coding with Discrete Constellations}
\label{sec:InfFin}

A second limitation of practical systems is  the use of suboptimal constellations with discrete points taken from a finite alphabet. In this case, perfect secrecy can still be achieved, but we must consider the constellation-constrained spectral efficiency $\hat{C}(\cdot)$ instead of $C(\cdot)$, i.e., (\ref{seccapa}) becomes
\begin{equation}
\label{eq19}
R_{n,k}(P_{n,k}) = \hat{C}(P_{n,k}H_{n,k}) - \hat{C}(P_{n,k}G_{n,k})\,.
\end{equation} 

In order to obtain simple resource allocation algorithms we consider the fitting of $\hat{C}(x)$ with a  linear combination of logarithmic functions, i.e.,
\begin{equation}
\label{eq20}
\hat{C}(x) \approx \beta_1 + \beta_2 \log(1 + \beta_3 x) - \beta_4 \log(1 + \beta_5 x)\,.
\end{equation}
In addition, when the \ac{SNR} $x$ is larger than a prefixed threshold, $\hat{C}(x)$ is clipped to the number of bits per symbol of the discrete constellation in order to model the saturation of the constellation-constrained spectral efficiency function.

\begin{figure}
\includegraphics[width=1\hsize]{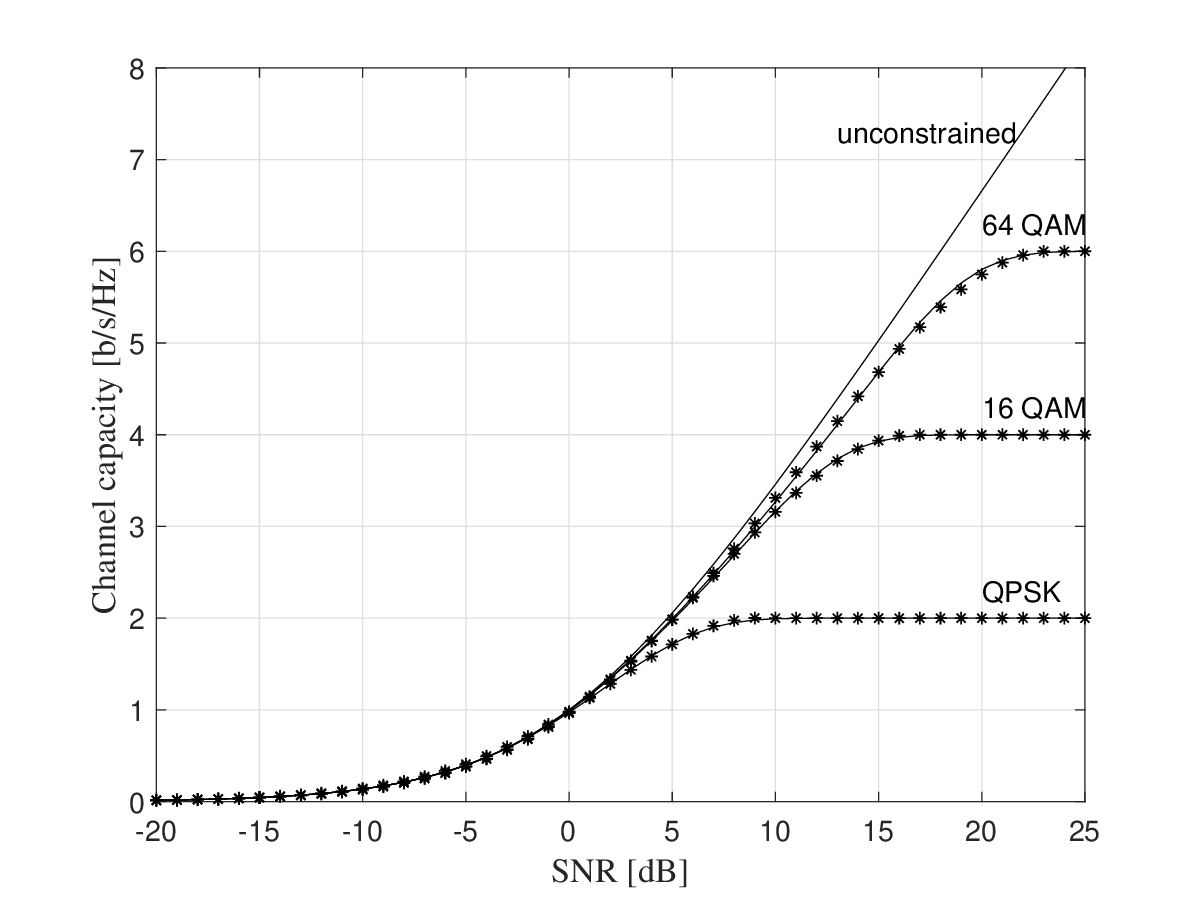}
\caption{The approximation (\ref{eq20}) of $\hat{C}(x)$ as a function of the \ac{SNR} (starred) and comparison with the exact spectral efficiency (continuous) for some QAM constellations.}
\label{fittingC}
\end{figure}

Fig. \ref{fittingC} shows the approximation  (\ref{eq20}) of $\hat{C}(x)$ as a function of the \ac{SNR} for QPSK, 16-QAM and 64-QAM constellations, and its comparison with the exact function. We observe a good agreement between the approximated and the exact curves.

\subsection{Finite-length Coding with Discrete Constellations}
\label{sec:finfin}

Let us consider the limitations introduced in Sections \ref{sec:FinInf} and \ref{sec:InfFin} jointly, i.e., both finite-length coding and discrete constellations, which describe a practical scenario. Also in this case we resort to the equivocation rate for the definition of the achievable secrecy rate (see problems (\ref{delRout1}) and (\ref{delRout2})), by replacing the spectral efficiency  $C(P)$ with the constellation-constrained spectral efficiency $\hat{C}(P)$ in (\ref{eq:eqbound}). On the other hand, since the approximation provided by \cite{Polyanskiy-13} is valid for any input distribution, \eqref{eq:polyanskiy} still holds true. 

\begin{figure}
\includegraphics[width=1\hsize]{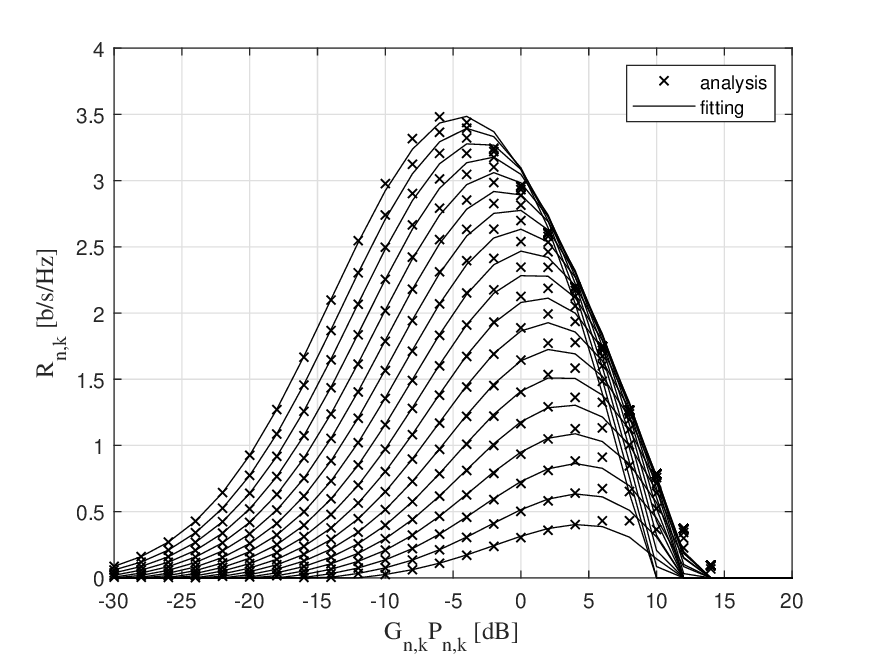}
\caption{$R_{n,k}(P_{n,k})$ as a function of $G_{n,k}P_{n,k}$ for values of $H_{n,k}/G_{n,k}$ between 2 dB and 20 dB with a step of 1 dB, and results obtained with the fitting function (\ref{fitfunc}), considering a $16$-QAM constellation and $m=4,096$.}
\label{finfin_fit}
\end{figure}

As already done in the previous section, we propose to fit $R_{n,k}(P_{n,k})$ by the function (\ref{fitfunc}). Fig. \ref{finfin_fit} shows $R_{n,k}(P_{n,k})$ for values of $H_{n,k}/G_{n,k}$ between 2 dB and 20 dB with a step of 1 dB, and results obtained by the fitting function (\ref{fitfunc}) with $\kappa = 10^{-3}$, 16-QAM constellation and $m = 4,096$.

\subsection{$\epsilon$-Outage Achievable Secrecy Rate}
\label{sec:epsilon}

In many practical scenarios Alice and the relays have only a partial \ac{CSI} of their channels to Eve. This is mainly due to the fact that Eve may not have an advantage in revealing its channels, e.g., by transmitting, unless this could be useful to increase the rate of other messages exchanged between her and the legitimate nodes. Indeed, in the absence of full \ac{CSI} there is a non-zero probability ({\rm outage probability})  that for any power allocation and choice of the secret message rate Eve may get some information on $\mathcal M$. 

In particular, we focus on the secrecy outage probability in each transmission phase and for each channel. Let  $\pi_{n,k}$ and $\bar{\pi}_{n,k}$ be the secrecy outage probabilities on channel $k$ with respect to relay $n$ in the first and the second phase, when messages are transmitted at rates $R_{n,k}(P_{n,k})$ and $\bar{R}_{n,k}(\bar{P}_{n,k})$, respectively. We consider as design criterion the limitation of the secrecy outage probability on each channel, i.e.,
\begin{equation}
\pi_{n,k} \leq \epsilon\,, \quad \bar{\pi}_{n,k} \leq \epsilon\,,
\label{outcond}
\end{equation}
where $\epsilon$ is the target secrecy outage probability. 

In the following we assume that the legitimate nodes know the statistics of both $G_{n,k}$ and $\bar{G}_{n,k}$, thus having a partial \ac{CSI}. If $R_{n,k}(P_{n,k})$ is the achievable secrecy rate for Alice-Eve channel realization $G^*_{n,k}$, then the secrecy outage probability can be written as 
\begin{equation}
\pi_{n,k}  = \mathbb{P}[G_{n,k} > G^*_{n,k}]\,.
\end{equation}
Similar expressions are obtained for the second phase. From (\ref{outcond}) we define $F_\epsilon$ as the {\em outage gain}, i.e., the channel gain for which 
\begin{equation}
\label{outgain}
\pi_{n,k} = \mathbb{P}[G_{n,k} > F_\epsilon] = \epsilon\,.
\end{equation}
Then the $\epsilon$-outage achievable secrecy rate can be obtained from the previous sections by considering $G_{n,k} = \bar{G}_{n,k} = F_\epsilon$.

\section{Single Link Power Optimization}
\label{sec:pointtopoint}

We first consider the single-link power optimization, where we allocate powers that maximize the secrecy sum rate between two nodes, using parallel channels. This problem must be solved in both transmission phases and here we focus on the first phase, i.e., the optimization of the communication from Alice to a specific relay $n$, assuming that all power $P_{\rm tot,1}$ can be used on that link. In this situation we have $P_{n',k} = 0$ for $n' \neq n$, $n=1, \ldots, N$, $k=1, \ldots, K$ and we must solve
\begin{equation}
\label{prob1}
R_{\rm max} = \max_{\{P_{n,k}\}} \; \sum_{k=1}^{K} R_{n,k}(P_{n,k})\,, \quad \mbox{s.t. (\ref{powerconst}).}
\end{equation}

The four cases of previous section are considered, i.e., $a)$ infinite-length coding with Gaussian constellations, $b)$ finite-length coding with Gaussian constellations, $c)$ infinite-length coding with discrete constellations, and $d)$ finite-length coding with discrete constellations. Moreover, we consider here the case of $\epsilon$-outage rates discussed in Section \ref{sec:epsilon}, thus considering gain $F_\epsilon$ for all channels to Eve.

\subsection{Infinite-length Coding with Gaussian Constellations}

For infinite-length coding with Gaussian constellations, the optimization problem (\ref{prob1}) has been solved in \cite{Jorswieck2008}. In particular, we immediately see that all channels for which $H_{n,k} < F_\epsilon$ must be switched off ($P_{n,k} = 0$) since they do not provide any secrecy rate. Let the set of used channels be 
\begin{equation}
{\mathcal F} = \{k: H_{n,k} > F_\epsilon\}\,.
\end{equation}
Then we have 
\begin{equation}
\begin{split}
P_{n,k} = 
& =\left[-\frac{H_{n,k} + F_\epsilon}{2 F_\epsilon  H_k } + \right. \\
& \left. \frac{\sqrt{(H_{n,k} - F_\epsilon)^2 + 4F H_{n,k} (H_{n,k}  - F_\epsilon)/\lambda)}}{2 F_\epsilon  H_{n,k} } \right]^+\,,
\end{split}
\label{solpk}
\end{equation}
where $\lambda$ is the Lagrange multiplier to be optimized in order to satisfy the power constraint, which can be computed by a dichotomic search.

\subsection{Finite-length Coding with Gaussian Constellations}

For finite-length coding with Gaussian constellations we exploit the fitting (\ref{fitfunc}) and the optimization problem (\ref{prob1}) becomes 
\begin{subequations}
\label{probfinite}
\begin{equation}
\begin{split}
R_{\rm max}& = \max_{\{P_{n,k}\}} \;  \sum_{k=1}^{K} \alpha_1 + \alpha_2 \log(1 + \alpha_3 H_{n,k} P_{n,k}) - \\
& \alpha_4 \log(1 + \alpha_5 H_{n,k} P_{n,k}) -  \alpha_6 \log(1 + \alpha_7 F_\epsilon P_{n,k}) + \\
& \alpha_8 \log(1 + \alpha_9 F_\epsilon P_{n,k})\,,
\end{split}
\end{equation}
\begin{equation}
\mbox{subject to (\ref{powerconst}).}
\end{equation}
\end{subequations}

By applying the Lagrange multipliers method we obtain
\begin{equation}
\label{eqlag}
\begin{split}
\frac{\alpha_2 \alpha_3 H_{n,k}}{1 + \alpha_3 H_{n,k} P_{n,k}} + \frac{\alpha_4 \alpha_5 H_{n,k}}{1 + \alpha_5 H_{n,k} P_{n,k}} + \\
+ \frac{\alpha_6 \alpha_7 F_\epsilon}{1 + \alpha_7 F_\epsilon P_{n,k}} +  \frac{\alpha_8 \alpha_9 F_\epsilon}{1 + \alpha_9 F_\epsilon P_{n,k}}  - \lambda = 0\,,
\end{split}
\end{equation}
where $\lambda$ is the Lagrange multiplier to be chosen in order to satisfy the power constraint.

Hence by defining
\begin{subequations}
\begin{equation}
\begin{split}
A_{n,k} & = \lambda \ln(2) \alpha_3 \alpha_5 \alpha_7 \alpha_9 H_{n,k}^2 F_{\epsilon}^2, \\
\end{split}
\end{equation}
\begin{equation}
\begin{split}
B_{n,k} & = \alpha_3 \alpha_5 \alpha_6 \alpha_7 \alpha_9 H_{n,k}^2 F_{\epsilon}^2 - \alpha_3 \alpha_5 \alpha_7 \alpha_8 \alpha_9 H_{n,k}^2 F_{\epsilon}^2 - \\
& \alpha_2 \alpha_3 \alpha_5 \alpha_7 \alpha_9 H_{n,k}^2 F_{\epsilon}^2 + \alpha_3 \alpha_4 \alpha_5 \alpha_7 \alpha_9 H_{n,k}^2 F_{\epsilon}^2 + \\
& \lambda \ln(2) \alpha_5 \alpha_7 \alpha_9 H_{n,k} F_{\epsilon}^2 + \lambda \ln(2) \alpha_3 \alpha_7 \alpha_9 H_{n,k} F_{\epsilon}^2 + \\
& \lambda \ln(2) \alpha_3 \alpha_5 \alpha_9 H_{n,k}^2 F_{\epsilon} + \lambda \ln(2) \alpha_3 \alpha_5 \alpha_7 H_{n,k}^2  F_{\epsilon}, \\
\end{split}
\end{equation}
\begin{equation}
\begin{split}
C_{n,k} & = \lambda \ln(2) \alpha_3 \alpha_5 H_{n,k}^2 - \alpha_2 \alpha_3 \alpha_5 \alpha_9 H_{n,k}^2 F_{\epsilon} + \\
& \lambda \ln(2) \alpha_7 \alpha_9 F_{\epsilon}^2 + \lambda \ln(2) \alpha_5 \alpha_7 H_{n,k} F_{\epsilon} - \\
& \alpha_2 \alpha_3 \alpha_5 \alpha_7 H_{n,k}^2 F_{\epsilon} + \alpha_4 \alpha_5 \alpha_7 \alpha_9 H_{n,k} F_{\epsilon}^2 + \\
& \alpha_3 \alpha_4 \alpha_5 \alpha_9 H_{n,k}^2 F_{\epsilon} + \lambda \ln(2) \alpha_3 \alpha_9 H_{n,k} F_{\epsilon} + \\
& \lambda \ln(2) \alpha_3 \alpha_7 H_{n,k} F_{\epsilon} + \alpha_3 \alpha_6 \alpha_7 \alpha_9 H_{n,k} F_{\epsilon}^2 + \\
& \alpha_3 \alpha_5 \alpha_6 \alpha_7 H_{n,k}^2 F_{\epsilon} - \alpha_5 \alpha_7 \alpha_8 \alpha_9 H_{n,k} F_{\epsilon}^2 - \\
& \alpha_3 \alpha_7 \alpha_8 \alpha_9 H_{n,k} F_{\epsilon}^2 - \alpha_3 \alpha_5 \alpha_8 \alpha_9 H_{n,k}^2 F_{\epsilon} + \\
& \lambda \ln(2) \alpha_5 \alpha_9 H_{n,k} F_{\epsilon} - \alpha_2 \alpha_3 \alpha_7 \alpha_9 H_{n,k} F_{\epsilon}^2 + \\
& \alpha_3 \alpha_4 \alpha_5 \alpha_7 H_{n,k}^2 F_{\epsilon} + \alpha_5 \alpha_6 \alpha_7 \alpha_9 H_{n,k} F_{\epsilon}^2, \\
\end{split}
\end{equation}
\begin{equation}
\begin{split}
D_{n,k} & = \alpha_4 \alpha_5 \alpha_7 H_{n,k} F_{\epsilon} - \alpha_2 \alpha_3 \alpha_9 H_{n,k} F_{\epsilon} - \\
& \alpha_5 \alpha_8 \alpha_9 H_{n,k} F_{\epsilon} - \alpha_3 \alpha_8 \alpha_9 H_{n,k} F_{\epsilon} + \\
& \alpha_5 \alpha_6 \alpha_7 H_{n,k} F_{\epsilon} - \alpha_2 \alpha_3 \alpha_7 H_{n,k} F_{\epsilon} + \\
& \alpha_4 \alpha_5 \alpha_9 H_{n,k} F_{\epsilon} - \alpha_2 \alpha_3 \alpha_5 H_{n,k}^2 + \\
& \alpha_3 \alpha_6 \alpha_7 H_{n,k} F_{\epsilon} + \alpha_6 \alpha_7 \alpha_9 F_{\epsilon}^2 - \alpha_7 \alpha_8 \alpha_9 F_{\epsilon}^2 + \\
& \lambda \ln(2) \alpha_9 F_{\epsilon} + \lambda \ln(2) \alpha_7 F_{\epsilon} + \lambda \ln(2) \alpha_5 H_{n,k} + \\
& \alpha_3 \alpha_4 \alpha_5 H_{n,k}^2 + \lambda \ln(2) \alpha_3 H_{n,k}, \\
\end{split}
\end{equation}
\begin{equation}
\begin{split}
E_{n,k} & = \lambda \ln(2) - \alpha_2 \alpha_3 H_{n,k} - \alpha_8 \alpha_9 F_{\epsilon} + \\
& \alpha_6 \alpha_7 F_{\epsilon} + \alpha_4 \alpha_5 H_{n,k}
\end{split}
\end{equation}
\label{eq:coeffs}
\end{subequations}
the Lagrangian (\ref{eqlag}) becomes 
\begin{equation}
\label{polyn}
\begin{split}
A_{n,k} P_{n,k}^4 + B_{n,k} P_{n,k}^3 + C_{n,k} P_{n,k}^2 \\ + D_{n,k} P_{n,k} + E_{n,k} = 0\,.
\end{split}
\end{equation}
For all real positive roots of the polynomial, we compute (\ref{eq19}) and select the root yielding the highest secrecy rate. When no real roots are found, it means that the secrecy rate is strictly decreasing for $P_{n,k} > 0$, thus $P_{n,k} = 0$ and a null secrecy rate is achieved.

Note that the algorithm must include a dichotomic search over $\lambda$ in order to satisfy the power constraints. Again, note that the solution to problem (\ref{probfinite}) is a generalization of the solution (\ref{solpk}) for ideal transmission conditions.

\subsection{Infinite-length Coding with Discrete Constellations}

For infinite-length coding with discrete constellations, the optimization problem (\ref{prob1}) using the fitting (\ref{eq20}) becomes
\begin{subequations}
\label{prob2}
\begin{equation}
\begin{split}
R_{\rm max}& = \max_{\{P_{n,k}\}} \;  \sum_{k=0}^{K-1} \left\{\beta_1 + \beta_2 \log(1 + \beta_3 H_{n,k} P_{n,k}) - \right.\\
&\beta_4 \log(1 + \beta_5 H_{n,k} P_{n,k})- \\
& \left. \left[ \beta_1 + \beta_2 \log(1 + \beta_3 F_\epsilon P_{n,k})\right. \right.\\
&\left.\left. - \beta_4 \log(1 + \beta_5 F_\epsilon P_{n,k})\right]\right\}^+\,,
\end{split}
\end{equation}
\begin{equation}
\mbox{subject to (\ref{powerconst}).}
\end{equation}
\end{subequations}

By comparing (\ref{probfinite}) with (\ref{prob2}), we note that the two problems are very similar, hence by applying also in this case the Lagrange multiplier method we obtain again (\ref{polyn}), with the following  coefficient values
\begin{equation} 
\begin{split}
\alpha_1  = 0, \quad 
\alpha_2  = \alpha_6 = \beta_2, \quad 
\alpha_3  = \alpha_7 = \beta_3, \\
\alpha_4  = \alpha_8 = \beta_4, \quad 
\alpha_5  = \alpha_9 = \beta_5. \\
\end{split}
\end{equation}

\subsection{Finite-length coding with discrete constellation}
For finite-length coding with discrete constellations, the optimization problem (\ref{prob1}) using the fitting (\ref{fitfunc}) becomes (\ref{probfinite}) and the  Lagrange multiplier methods leads to \eqref{eq:coeffs}.

\section{Maximum Rate Power Allocation}
\label{sec:MRallocSec}

We now consider the power allocation problem at Alice and Bob with the aim of maximizing the secrecy rate, i.e.,
\begin{subequations}
\label{optprob}
\begin{equation}
\label{subopt1}
R_{\rm max} = \max_{\bm{P}, \bar{\bm{P}}} R_{\rm tot}(\bm{P}, \bar{\bm{P}})\,,
\end{equation}
\begin{equation}
\label{subopt2}
\mbox{subject to power constraints (\ref{powerconst}),}
\end{equation}
\begin{equation}
\mbox{and rate constraints (\ref{ratecons1}) and (\ref{ratecons2}).}
\end{equation}
\end{subequations}

As observed in \cite{Tomasin2015}, this is a mixed-integer programming problem, and for its solution we resort to the iterative approach of \cite{Tomasin2015}  based on  the game-theoretic  Gale and Shapley algorithm for the stable matching problem \cite{Shapley-62}. Next we summarize the algorithm, while referring the reader to  \cite{Tomasin2015} for its detailed description.

The stable matching problem aims at matching dames to cavaliers, without having a  dame and a cavalier belonging to two different couples both preferring to be matched. In our scenario, dames and cavaliers are channels and relay, respectively, and the preference of matching is the achievable secrecy rate when using the channel for that relay. We have actually two coupled stable matching problems for the two phases. We use an iterative algorithm, where at each iteration one step of the Gale and Shapley algorithm is performed for both problems.  

We start computing the overall rates obtained by assigning all channels to each relay in phase 2 (finding the best power allocation for both phases and the best channel assignment in phase 1), and then we exclude the  relay-channel couple in phase 2 that provides the lowest rate. At the second iteration we compute the overall rates obtained by assigning all channels (except the couple excluded in the first iteration) to each relay in phase 2 (again optimizing powers and phase-1 channel allocation), before excluding another  relay-channel couple  in phase 2 that provides the lowest rate. The process is iterated excluding a couple at each iteration until for each channel we have at most one associated relay in phase 2. Within each iteration the channel allocation for phase 1 is obtained again by the Gale and Shapley algorithm applied to the matching of channels and relays in phase 1 (for a given allocation in phase 2).

\section{Numerical Results}
\label{sec:num_results}

\subsection{Simulation Scenario}

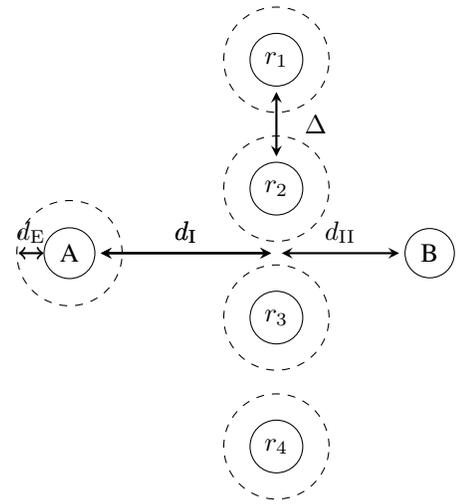
\begin{figure}
\centering
\scalebox{1}{
\begin{tikzpicture}
	\node[circle,draw] (tx) {A};
	
	% Alice-relay channel	
	\node[coordinate, right= of tx, xshift=4em] (centro) {};
	\node[circle, draw, right= of centro, xshift=2em] (rx) {B};
	\node[circle, draw, above= of centro, yshift=-1.42em] (r2) {$r_2$};
    
	\node[circle, draw, above= of r2] (r1) {$r_1$};
	\node[circle, draw, below= of r2] (r3) {$r_3$};	
	\node[circle, draw, below= of r3] (r4) {$r_4$};	
	
    \draw[dashed] (tx) circle (0.7cm);%  -- +  (0:-1cm)  node(B){z};
 	\draw[dashed] (r1) circle (0.7cm);%  -- +  (0:-1cm)  node(B){z};
	\draw[dashed] (r2) circle (0.7cm);%  -- +  (0:-1cm)  node(B){z};
	\draw[dashed] (r3) circle (0.7cm);%  -- +  (0:-1cm)  node(B){z};
	\draw[dashed] (r4) circle (0.7cm);%  -- +  (0:-1cm)  node(B){z};

	\node(B) at ([shift={(tx)}] 0:-0.8cm) {};
	
	\draw[thick,<->,shorten >=2pt,shorten <=2pt,>=stealth] (tx) -- node[above]   {$d_{\rm I}$} (centro) ;	
	\draw[thick,<->,shorten >=2pt,shorten <=2pt,>=stealth] (centro) -- node[above] {$d_{\rm II}$}(rx) ;	
	\draw[thick,<->,shorten >=2pt,shorten <=2pt,>=stealth] (r1) -- node[right,xshift=0.7em] {$\Delta$}(r2) ;	
 	\draw[thick,<->,shorten >=2pt,shorten <=2pt,>=stealth] (tx) -- node[above]   {$d_{\rm I}$} (centro) ;	
	\draw[thick,<->] (tx) -- node[above]   {$d_{\rm E}$} (B) ;

\end{tikzpicture}}
\caption{Node position diagram. A: Alice, B: Bob, $r_n$: relay $n$.}
\label{nodepos}
\end{figure}

Let us consider the scenario reported in Fig. \ref{nodepos}, where the relay nodes are positioned along a line that is orthogonal to the segment between Alice to Bob, intersecting it at a distance $d_{\rm I}$ and $d_{\rm II}$ from Alice and Bob, respectively. Moreover, relays are equispaced with a distance $\Delta$ between any two adjacent relays. We further assume that the eavesdropper is at least at a distance $d_{\rm E}$ from any transmitting node, i.e., it is outside of the dashed circles surrounding Alice and the relays.

The $K=16$ channels between any couple of nodes are assumed independent Rayleigh fading. We also consider $P_{\rm tot,1} = P_{\rm tot,2} = 1$. The average \ac{SNR} at unitary distance is of 0 dB, and the path loss coefficient is  3.5, thus the average \ac{SNR} at distance $d$ is $d^{-3.5}$. About the eavesdropper, since it is assumed to be at a minimum distance $d_{\rm E}$ from any transmitting node, the outage gain is obtained from (\ref{outgain}) and from the Rayleigh fading assumption  $\mathbb{P}[G_{n,k} \leq F_\epsilon] = \exp[-F_\epsilon/(d_{\rm E}^{-3.5})]$, therefore
\begin{equation}
F_\epsilon =  - d_{\rm E}^{-3.5} \ln \epsilon\,.
\end{equation}
For finite-length coding we assume a \ac{CER} at Bob $\kappa = 10^{-3}$ and $m = 128$ and $4,096$.

\begin{figure}
\includegraphics[width=1\hsize]{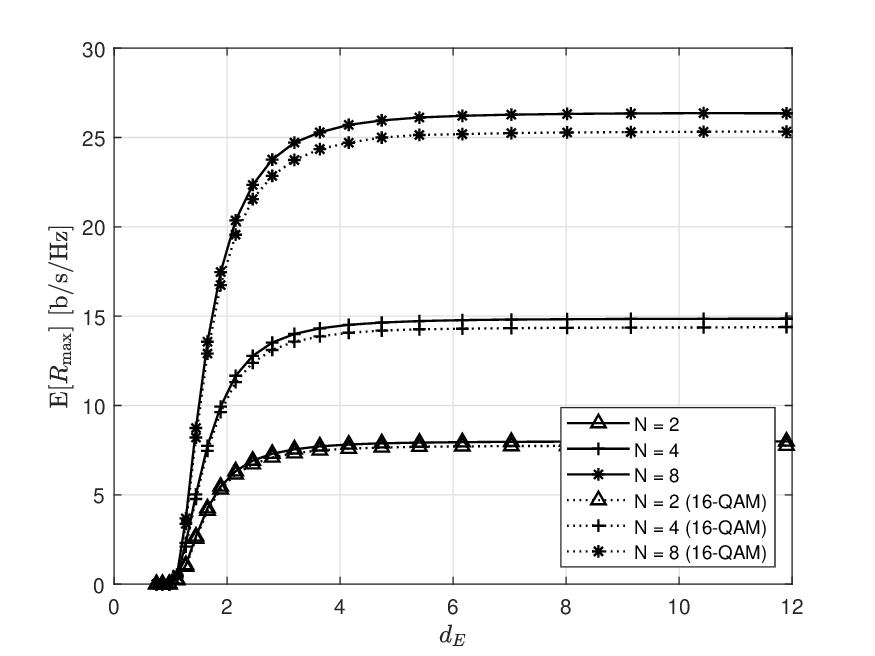}
\caption{Average maximum secrecy rate as a function of $d_{\rm E}$ with infinite-length coding, both Gaussian and discrete (16-QAM) constellations and various values of $N$.}
\label{figthr_infcod}
\end{figure}

\begin{figure}
\includegraphics[width=1\hsize]{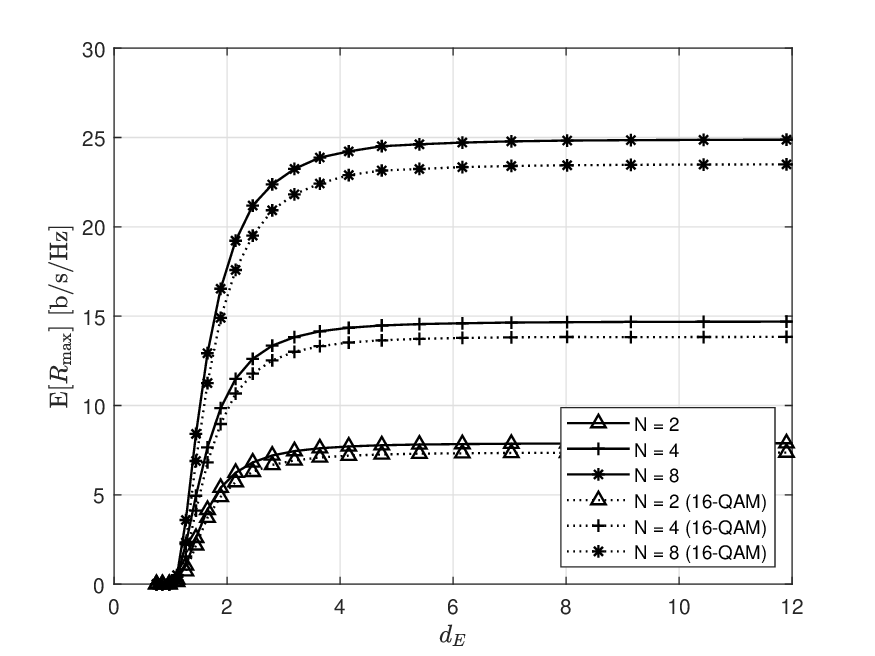}
\caption{Average maximum outage secrecy rate as a function of $d_{\rm E}$ with finite-length coding ($ m = 4,096$), both Gaussian and discrete (16-QAM) constellations and various values of $N$.}
\label{finlen_bothcases}
\end{figure}

\begin{figure}
\includegraphics[width=1\hsize]{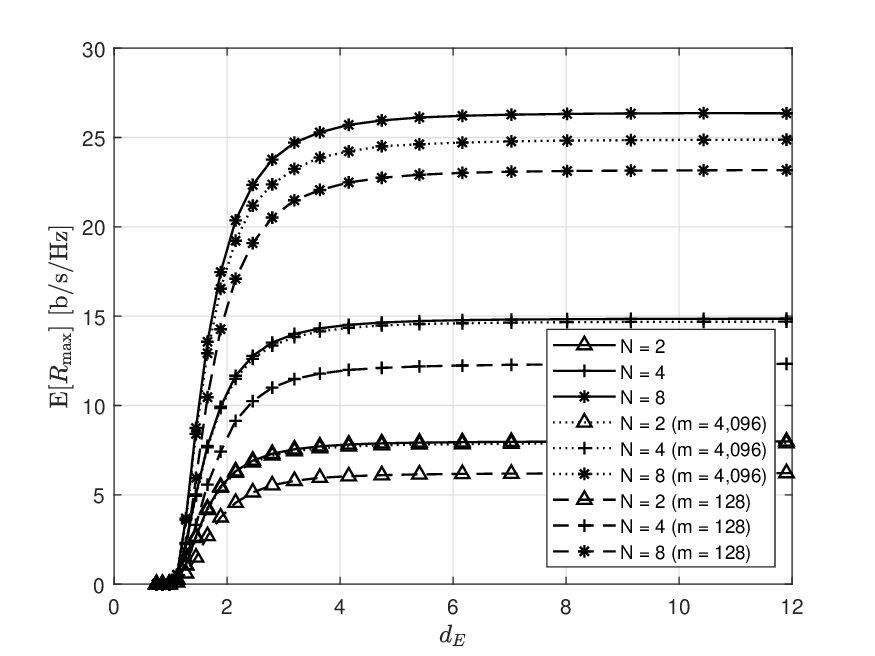}
\caption{Average maximum outage secrecy rate as a function of $d_{\rm E}$ with Gaussian constellation, both infinite- and finite-length ($m=128$ and $m=4,096$)  coding  and various values of $N$.}
\label{contmod_bothcases}
\end{figure}

\begin{figure}
\includegraphics[width=1\hsize]{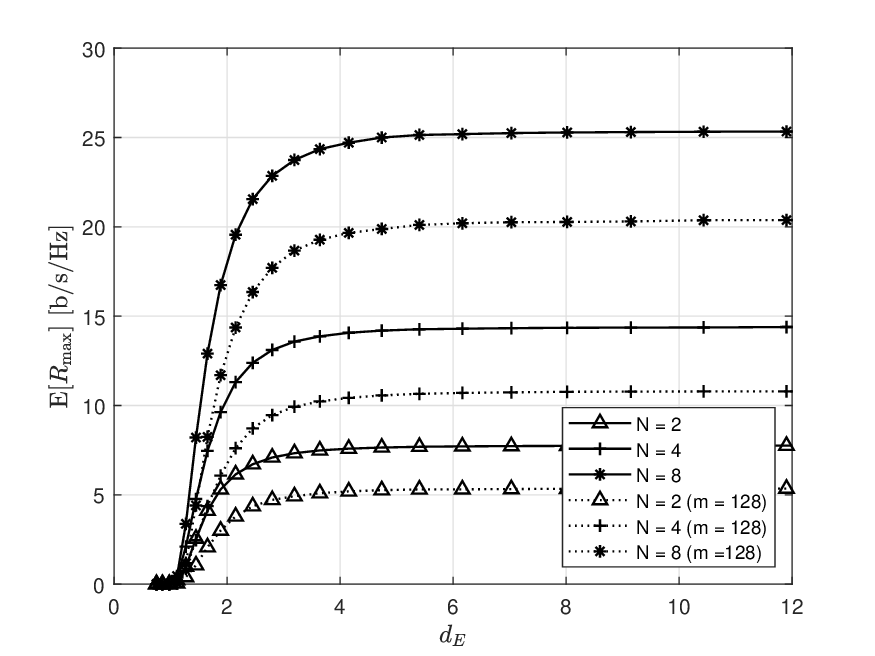}
\caption{Average maximum outage secrecy rate as a function of $d_{\rm E}$ with 16-QAM, both infinite and finite-length coding ($m=128$) and various values of $N$.}
\label{figthr_finmod_QAM}
\end{figure}

\subsection{Impact of Eve's distance}

We first consider a scenario in which each relay has the same distance from Alice and Bob, i.e., $d_{\rm I} = d_{\rm II} = 0.8$, the separation between relays is $\Delta = 0.05$, the number of relays is $N=2$, 4 or 8. 

Figs. \ref{figthr_infcod}-\ref{figthr_finmod_QAM} show the average maximum outage secrecy rate $\mathbb{E}[R_{\rm max}]$, averaged over channel realizations, as a function of $d_{\rm E}$, for a target secrecy outage probability $\epsilon = 10^{-4}$, and comparing different coding and constellations settings. In particular, Fig. \ref{figthr_infcod} reports results for a transmission using infinite-length coding and both Gaussian and discrete constellations, Fig. \ref{finlen_bothcases} shows results for finite-length coding and both Gaussian and discrete constellations. In both cases we note that, as expected, Gaussian signaling outperforms discrete modulation (16-QAM) in terms of secrecy rate, regardless of the use of infinite- or finite-length codes. Moreover, by increasing the number of relays, the average maximum outage secrecy rate increases, as a diversity gain is available on the links among legitimate nodes. Moreover, as $d_{\rm E} \rightarrow \infty$ we note that the rate curves flattens in correspondence of the unsecure rate of the relay parallel channels, as in this case security conditions are always met and the performance is limited only by the legitimate channel conditions.

\subsection{Impact of Codeword Length}

In Figs. \ref{contmod_bothcases} and \ref{figthr_finmod_QAM} the impact of finite-length coding for both   Gaussian and discrete constellations is investigated. Note that the performance of codes with long codewords  ($m = 4,096$) is  comparable to that of infinite-length coding. Considering a 16-QAM, differences between infinite-length coding and 4,096-length coding are negligible as the average maximum outage secrecy rates coincide for all numbers of relay nodes. In Fig. \ref{figthr_finmod_QAM} we note that short codes ($m=128$) instead visibly degrade the average maximum secrecy rate.

\begin{figure}
\includegraphics[width=1\hsize]{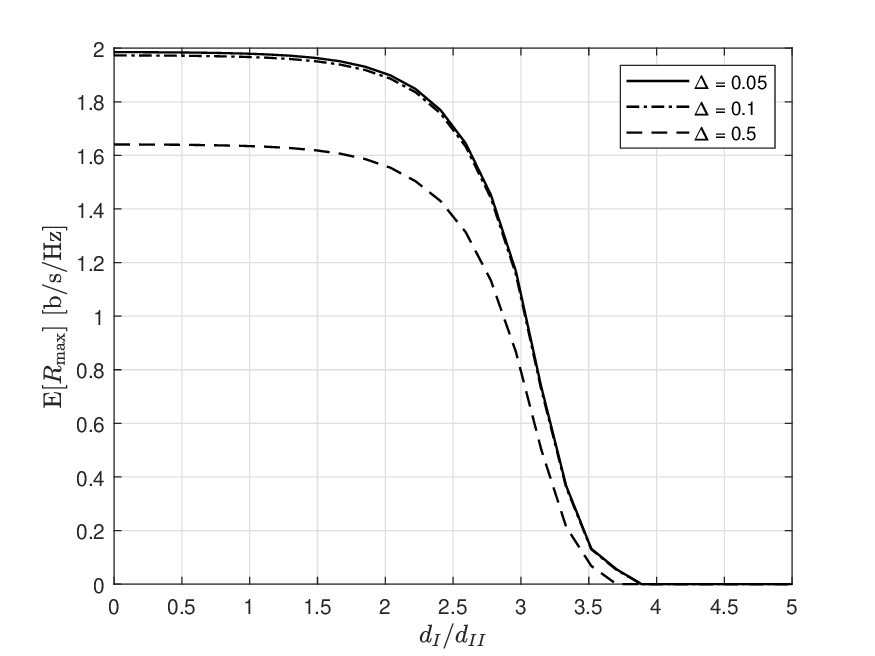}
\caption{Average maximum outage secrecy rate as a function of $d_{\rm I}/d_{\rm II}$, various values of $\Delta$, and finite-length coding ($m= 4,096$) with discrete (16-QAM) constellations, for $d_{\rm E} = 10$, $\epsilon = 10^{-4}$ and $N=4$ relays.}
\label{figthr_dIdII}
\end{figure}

\subsection{Impact of Relative Node Distances}

We then study the impact of the relative distances among  legitimate nodes. In particular, we fix the Alice-Bob distance to $d_{\rm I} + d_{\rm II} = 2$, and we let the ratio between the two distances $d_{\rm I}/d_{\rm II}$ and the distance among the relays vary, i.e., $\Delta = \{0.05, 0.1, 0.5\}$, for $d_{\rm E} = 10$, $\epsilon = 10^{-4}$ and $N=4$ relays. 

Fig. \ref{figthr_dIdII} shows the average maximum outage secrecy rate as a function of $d_{\rm I}/d_{\rm II}$, and finite-length coding ($m = 4,096$) with Gaussian constellations. We observe that for decreasing values of $\Delta$ the curves tend asymptotically to a maximum average secrecy rate of 2 b/s/Hz. On the other hand, as $d_{\rm I}/d_{\rm II}$ tends to infinity, the average maximum outage secrecy rate tends to zero, as the Alice-relay links will provide vanishing data rates. When the distance $\Delta$ tends to zero all the relay nodes are squeezed in the same point between Alice and Bob, which represents the optimal relaying configuration.

\subsection{Comparison With Other Solutions}

\begin{figure}
\includegraphics[width=1\hsize]{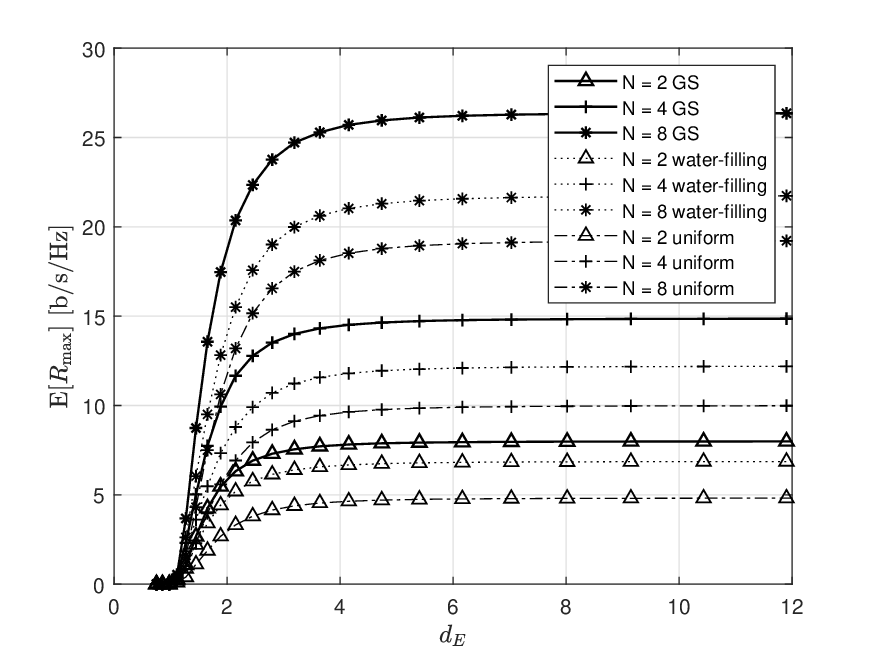}
\caption{Average maximum outage secrecy rate as a function of $d_E$, with infinite-length coding ($m= 4,096$) and Gaussian constellations, obtained using: GS power allocation, water-filling and uniform power allocation.}
\label{waterfill}
\end{figure}

\begin{figure}
\includegraphics[width=1\hsize]{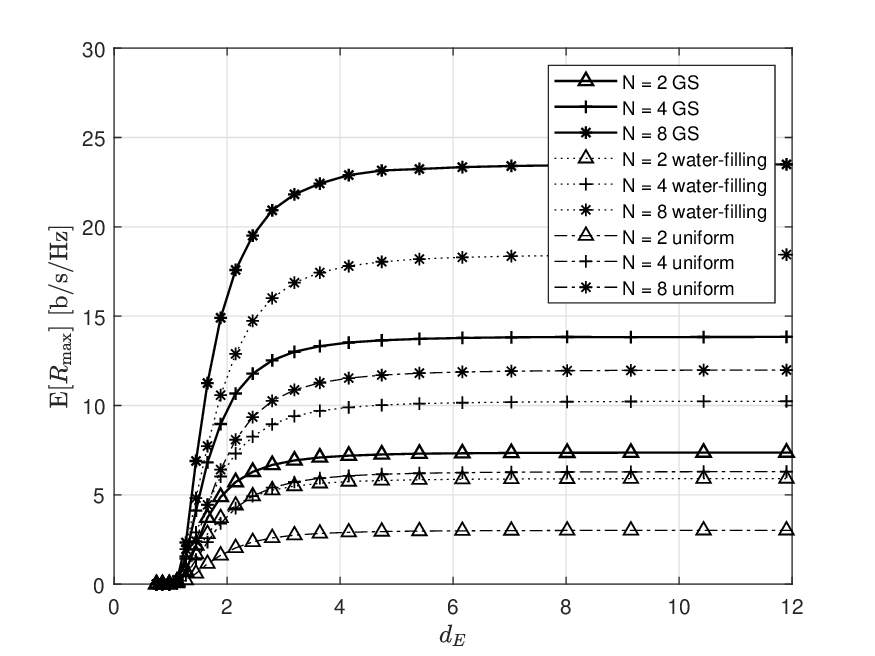}
\caption{Average maximum outage secrecy rate as a function of $d_E$, with finite-length coding ($m= 4,096$) and discrete (16-QAM) constellations, obtained using: optimal power allocation, water-filling and uniform power allocation..}
\label{waterfill_fin}
\end{figure}

Figs. \ref{waterfill} and \ref{waterfill_fin} provide a comparison between our resource allocation (denoted as Gale-Shapley, or GS) strategy and two suboptimal solutions, respectively uniform power allocation over the $K$ channels and water-filling allocation. Various scenarios are considered, i.e. infinite-length codes with Gaussian signaling and finite-length coding with discrete constellations. We also consider that each relay has the same distance from Alice and Bob, i.e., $d_{\rm I} = d_{\rm II} = 0.8$, the separation between relays is $\Delta = 0.05$, the number of relays is $N=2$, 4 or 8. Water-filling provides the best possible power allocation in Eve's absence, since it assigns more power to the channels presenting better gains. However, this solution  is not convenient from a security standpoint, since channels that are good for the legitimate receiver could also be good for the attacker, thus degrading the secrecy performance. As predictable, uniform allocation leads to the worst average secrecy rate for all the considered cases. 

\section{Conclusions}
\label{sec:conclusions}

In this paper we have derived the secrecy rate of the Gaussian relay parallel channel under finite-length coding and discrete constellation constraints, defined as the maximum rate for which a minimum equivocation rate is achieved at Eve. Moreover, we have applied a coupled version of the Gale and Shapley algorithm to allocate power within each channel in order to maximize the secrecy rate. Numerical results show the effectiveness of the resource allocation approach we consider, and show that moderate sizes of both the constellation alphabet and the codewords are sufficient to achieve close-to-optimal secrecy rates for typical wireless transmission scenarios.

\bibliographystyle{IEEEtran}
\bibliography{biblio}

\end{document}